\documentclass[sigconf]{acmart}

\usepackage{float}
\usepackage{enumitem}
\usepackage{makecell}

\usepackage{multirow}

\def \basp {-.3ex}
\def \insp {-.3ex}

\newcommand{\be}{ \begin{enumerate}[topsep=\basp,itemsep=0pt,partopsep=0pt, parsep=\insp]}

\newcommand{\ee}{\end{enumerate}}
\AtBeginDocument{%
  \providecommand\BibTeX{{%
    \normalfont B\kern-0.5em{\scshape i\kern-0.25em b}\kern-0.8em\TeX}}}

\copyrightyear{2024}
\acmYear{2024}
\setcopyright{acmlicensed}\acmConference[ASSETS '24]{The 26th International ACM SIGACCESS Conference on Computers and Accessibility}{October 27--30, 2024}{St. John's, NL, Canada}
\acmBooktitle{The 26th International ACM SIGACCESS Conference on Computers and Accessibility (ASSETS '24), October 27--30, 2024, St. John's, NL, Canada}
\acmDOI{10.1145/3663548.3675659}
\acmISBN{979-8-4007-0677-6/24/10}

\begin{document}

\title{Misfitting With AI: How Blind People Verify and Contest AI Errors}


\author{Rahaf Alharbi}
\email{rmalharb@umich.edu}
\affiliation{%
  \institution{University of Michigan}
  \city{Ann Arbor}
  \country{USA}
}

\author{Pa Lor}
\email{palor@umich.edu}
\affiliation{%
  \institution{University of Michigan}
  \city{Ann Arbor}
  \country{USA}
}

\author{Jaylin Herskovitz}
\email{jayhersk@umich.edu}
\affiliation{%
  \institution{University of Michigan}
  \city{Ann Arbor}
  \country{USA}
}

\author{Sarita Schoenebeck}
\email{yardi@umich.edu}
\affiliation{%
  \institution{University of Michigan}
  \city{Ann Arbor}
  \country{USA}
}

\author{Robin Brewer}
\email{rnbrew@umich.edu}
\affiliation{%
  \institution{University of Michigan}
  \city{Ann Arbor}
  \country{USA}
}

\renewcommand{\shortauthors}{Alharbi et al.}

\begin{abstract}
Blind people use artificial intelligence-enabled visual assistance technologies (AI VAT) to gain visual access in their everyday lives, but these technologies are embedded with errors that may be difficult to verify non-visually. Previous studies have primarily explored sighted users' understanding of AI output and created vision-dependent explainable AI (XAI) features. We extend this body of literature by conducting an in-depth qualitative study with 26 blind people to understand their verification experiences and preferences. We begin by describing errors blind people encounter, highlighting how AI VAT fails to support complex document layouts, diverse languages, and cultural artifacts. We then illuminate how blind people make sense of AI through experimenting with AI VAT, employing non-visual skills, strategically including sighted people, and cross-referencing with other devices. Participants provided detailed opportunities for designing accessible XAI, such as affordances to support contestation. Informed by disability studies framework of misfitting and fitting, we unpacked harmful assumptions with AI VAT, underscoring the importance of celebrating disabled ways of knowing. Lastly, we offer practical takeaways for Responsible AI practice to push the field of accessible XAI forward.
 
\end{abstract}

\ccsdesc[500]{Human-centered computing~Empirical studies in accessibility}

\keywords{Accessibility, Visual Assistance Technology, Blind people, Verification, Artificial Intelligence, Explainability; Seeing AI; Be My Eyes}


\maketitle

\section{Introduction}
Artificial intelligence (AI) is pervasive in our everyday lives and routines. Computer vision, a type of AI, is integrated into various real-world systems, from social media filters to traffic cameras to medical imaging. Being able to understand, verify, or even contest AI systems has become consequential for myriad activities, ranging from routine interactions to high-stakes decision-making. For example, when posting images of people on Twitter (now X), users visually noticed that the cropping algorithm would only highlight White people, revealing a kind of racial bias \cite{nbc_twitter_investigate_bias, guardian_twitter_apology}. This incident, along with many other cases of users detecting AI limitations and harms \cite{google_apology, forbes_yelp_unbiased,vox_youtube_demonetization}, motivated a wealth of prior research to study how users make sense of AI \cite{shen2021everyday,eslami2019user, kim2023humans}. It also pushed Responsible AI researchers, regulators, and activists to advocate for explainable AI (XAI) as a means to support users' decision-making \cite{madiega2021artificial, panigutti2023role, almada2019human, lyons2021conceptualising}.

 However, understanding computer vision outputs often relies on a person's visual ability \cite{kim2023humans, alharbi2019mask, stock2018convnets}. As a result, much of the literature on XAI has focused on \textit{sighted} experiences and has designed vision-centered XAI measures such as displaying heatmaps to users \cite{maehigashi2023experimental, shitole2021one, kim2023humans}. Very little is known about how blind people non-visually make sense of AI outputs, and what types of XAI features or otherwise would be useful \cite{hong2022blind,macleod2017understanding}. This is especially crucial since error detection could be inaccessible to blind people \cite{macleod2017understanding, gonzalez2024investigating, alharbi2022understanding}, leading to the overtrust \cite{macleod2017understanding} or abandonment \cite{phillips1993predictors, kintsch2002framework} of AI systems. Thus, we pose and explore the following research questions: how do blind people verify AI results, and what (if any) types of features may support accessible verification?

 We focused on AI-enabled visual assistance technologies (AI VAT), which are real-world mobile applications that blind people use daily to gain visual access, from reading to cooking to describing scenes \cite{herskovitz2023hacking, ramil2021allergic,gonzalez2024investigating, stangl2020person}. In essence, we chose AI VAT to capture blind people's \textit{everyday} experience given its ubiquity in blind communities. Furthermore, blind people are concerned about the lack of accuracy in AI VAT, and the accompanying risks of wrong outputs \cite{kim2022user,abdolrahmani2017embracing,gonzalez2024investigating}. For instance, when using AI VAT to navigate an unfamiliar space, blind people worry about stigmatizing errors such as mislabeled bathroom signs \cite{abdolrahmani2017embracing}. While certain XAI features (e.g., conveying confidence ratings \cite{macleod2017understanding}) may support blind people in negotiating errors, these features have not been included in commercially available AI VAT.
 
 We conducted semi-structured interviews with 26 blind people who frequently use AI VAT. Our analysis revealed common errors blind people face when using AI VAT, highlighting processing limitations and cross-cultural bias. Then, we uncovered how blind people verify AI VAT by employing various methods such as: testing AI VAT in low-risk and familiar contexts, engaging non-visual sensemaking skills, strategically including sighted people, and switching between various applications and devices. Lastly, participants outlined opportunities for accessible XAI to support their verification work. They emphasized improved camera guidance within AI VAT, and they had mixed opinions about the added value of confidence ratings. Whether blind people perceived confidence ratings as positive or negative hinged upon the nature of how they used AI VAT.  While confidence ratings were sometimes perceived as a way to save time by allowing people to quickly disregard low-accuracy results, other participants felt that the effort it would take to parse this additional information would interrupt their workflow. Participants also desired ways to challenge and refine AI VAT results through direct feedback.

Informed by feminist disability studies \cite{Thomson1994-THORTB, garland2002integrating, shildrick1998rejected}, we used the concepts of misfitting and fitting \cite{garland2011misfits} to interpret our findings. Rosemarie Garland-Thomson developed the misfitting framework to describe the mutual and ever-shifting entanglements between bodies and environments. Misfitting and fitting ``denote an encounter in which two things come together in either harmony or disjunction'' \cite{garland2011misfits}. In the context of technology, AI systems are often built for dominant groups, casting those who fall outside of these groups as misfits. We argue that blind people ``misfit'' in computer vision systems as those technologies are predominately built for sighted people, using sighted data \cite{massiceti2023explaining, gurari2018vizwiz}. Blind people who hold marginalized identities experience an additional layer of misfitting when AI VAT fails to attend to their languages and cultural artifacts. Misfitting as an analytical lens positions disabled people as knowers and experts through their lived experience of navigating access. We highlight blind people's epistemic work of verifying AI VAT, calling for future technologies to celebrate and incorporate disabled ways of knowing.

This study has three main contributions. First, we present empirical data on the everyday experience of blind people who use AI VAT. We add to the growing body of literature on AI for visual access \cite{bennett2020care, gonzalez2024investigating, alharbi2022understanding, herskovitz2023hacking, stangl2020person, gamage2023blind, vincenzi2021interdependence} by outlining errors faced by blind people, illuminating how they verify AI outcomes, and their preferred roles in improving AI workflows. Our analysis contributes a holistic understanding of blind people's embodied and continuous process of interpreting AI and centers blind people's expectations for upcoming features that could support their existing work. Second, we extend the framework of misfitting and fitting \cite{garland2011misfits} in the context of assistive technologies to unearth the limitations of AI VAT. Third, we offer specific directions for Responsible AI grounded in participants' desires for contestable and accessible XAI.

\section{Related Work}
We begin by reviewing the intersection of feminist disability studies and assistive technology research, a grounding sensibility in our study. Then, we highlight prior work on visual assistance technologies, focusing on their affordances and limitations. Lastly, we synthesize the emerging area of XAI and its focus on understanding how end-users interpret AI, noting opportunities for accessibility research.

\subsection{Feminist Disability Studies \& Accessibility} 
Disability studies is an interdisciplinary and multidisciplinary field that broadly aims to examine the historical, political, and cultural contexts of disability \cite{bell2010disability, minich2016enabling, ferguson2012disability}. Recently, disability studies have become a central sensibility to many HCI and accessibility works (e.g., \cite{mankoff2010disability, hofmann2020living, mcdonnell2023easier, alharbi2022understanding}). In their foundational paper, Mankoff et al. argued that disability studies serve as a critical lens to complicate assistive technology research and identify harms, calling on researchers to engage more deeply with disabled people and move away from medical perspectives \cite{mankoff2010disability}. In 1994, disability studies scholar Rosemarie Garland-Thomson coined the term \textit{feminist disability studies} \cite{Thomson1994-THORTB}, emphasizing that disability is integral to feminist scholarship \cite{garland2002integrating}. Contrary to common misconceptions, feminist disability studies do not merely concern the experiences of disabled women. Rather, it provides an analytical framework to examine structures that construct disability \cite{garland2002integrating}, attending to questions of fluidity \cite{shildrick1998rejected} and knowledge making \cite{garland2011misfits, haraway2016situated}. 

Accessibility researchers similarly began incorporating feminist disability studies. For instance, Hofmann et al. built upon Mankoff et al. \cite{mankoff2010disability} and feminist stances \cite{haraway2016situated, suchman2020agencies} to advocate for 1) acknowledging and challenging ableism\footnote{Talila A. Lewis, community lawyer and organizer, defines ableism as ``[a] system of assigning value to people's bodies and minds based on societally constructed ideas of normalcy, productivity, desirability, intelligence, excellence, and fitness. These constructed ideas are deeply rooted in eugenics, anti-Blackness, misogyny, colonialism, imperialism, and capitalism. This systemic oppression that leads to people and society determining people's value based on their culture, age, language, appearance, religion, birth or living place, `health/wellness', and/or their ability to satisfactorily re/produce, `excel' and `behave.' You do not have to be disabled to experience ableism'' \cite{lewis2022working}.}, 2) recognizing the complexity of disability, and 3) grounding accessibility research in critical disability studies \cite{hofmann2020living}. Bennett et al. drew from feminist disability studies and science, technology, and society (STS) scholarship to attend to the care work of access, reorienting AI assistive technology research from an individualistic focus (i.e., what can a disabled person do) to a more collective approach \cite{bennett2020care}. Grounded in crip technoscience -- a feminist STS and crip theory approach developed by Hamraie and Fritsch to center the expertise of disabled people in the design process \cite{hamraie2019crip} -- Hsueh et al. reimagined accessible data visualizations by affirming how access is a continual and transformative process \cite{hsueh2023cripping}. 

We contribute to the growing area of feminist disability studies in accessibility by applying Rosemarie Garland-Thomson's framework of misfitting \& fitting \cite{garland2011misfits} to AI-based assistive technology. Situated in conversations with feminist scholarship on matters of care and dependency \cite{fineman2010vulnerable,kittay1999loves}, misfitting and fitting offers a materialist extension of the social model of disability \cite{scully2008disability, siebers2008disability} by directing attention to dynamic and mutually constructive relationships between bodies and worlds. Garland-Thomson explained ``the degree to which that shared material world sustains the particularities of our embodied life at any given moment or place determines our fit or misfit'' \cite{garland2011misfits}. Misfitting engenders discomfort and incompatibility whereas fitting entails comfort and harmony. People who fit are typically considered ``uniform, standard, majority bodies'' \cite{garland2011misfits}.  For Garland-Thomson, misfitting inspires expertise or crip wisdom \cite{williamson2019accessible, hsueh2023cripping, SinsInvalidAutostraddle,williams2021articulations}, highlighting the ways in which disabled people creatively subvert access barriers. Our paper uses misfitting to examine assistive and access technologies. Recently, Williams drew upon the misfitting framework to critique robotic interventions that are built to manage the "misfit" of autistic people, arguing for transformative futures that allow for solidarity \cite{williams2021misfit}. In the context of AI VAT, we question who misfits and fits in AI systems, noting how there is a stereotypical representation of a blind person that simultaneously fits and misfits in AI VAT.

\subsection{Visual Assistance Technologies (VAT)}
Broadly, visual assistance technologies (VAT) are camera-based mobile applications that add to blind people's existing ecosystem of assistive tools (e.g., adding braille stickers and engaging their mobility skills \cite{shinohara2007observing, blindness-identification}). VAT is generally split into two categories based on how visual information is provided: either from human assistance or AI systems. Human-enabled VAT applications can be mediated by volunteers (e.g., Be My Eyes\footnote{Be My Eyes: \href{https://www.bemyeyes.com/}{https://www.bemyeyes.com/}}), trained agents (e.g., Aira\footnote{Aira: \href{https://visualinterpreting.com/}{https://visualinterpreting.com/}}), or crowdworkers (e.g., VizWiz \cite{bigham2010vizwiz}). AI VAT, such as Microsoft Seeing AI\footnote{Seeing AI: \href{https://www.seeingai.com/}{https://www.seeingai.com/}}, TapTapSee\footnote{TapTapSee:\href{https://taptapseeapp.com/}{https://taptapseeapp.com/}}, Envision AI\footnote{Envision AI:\href{https://www.letsenvision.com/}{https://www.letsenvision.com/}}, and KNFB Reader\footnote{Now known as OneStep Reader: \href{https://sensotec.be/en/product/onestep-reader/}{https://sensotec.be/en/product/onestep-reader/}} \cite{knfbreader}, typically use techniques such as Optical Character Recognition (OCR) to recognize text in scanned documents, object recognition to identify products, scene description, color recognition, or facial recognition. Most recently,  Be My Eyes (a human-enabled VAT) introduced Be My AI, a feature that incorporates the large language model (LLM) GPT-4 developed by OpenAI, enabling blind people to submit images, receive visual descriptions, and ask follow-up questions \cite{BeMyAI}. Prior work has studied the appropriate use cases for human-enabled VAT and AI VAT \cite{kim2022user,gurari2017crowdverge,gurari2018vizwiz,brady2013visual}. Overall, they found that human-enabled VAT is suited for `subjective' tasks (e.g., receiving fashion advice \cite{burton2012crowdsourcing}) and complex tasks (e.g., obtaining the license plate of a ride share before entering the car \cite{brewer2019understanding}). In contrast, AI VAT is typically used for `objective' tasks such as reading \cite{neat2019scene}.
Although human-enabled VAT is currently more powerful, connecting to a human assistant can be time-consuming or costly \cite{avila2016remote,ferris2022beyond}. Thus, AI VAT remains an important tool for bolstering visual access.

However, errors are common when using AI VAT \cite{herskovitz2023hacking, kim2022user}. Partly, this may be because foundational AI models are built without direct engagement with blind communities. In their recent review of smart assistive technologies for visual access, Gamage et al. found that 82\% of studies did not involve blind people and that there is a disconnect between tasks that blind people wanted AI support with (e.g., completing paper forms) and the tasks researchers prioritized \cite{gamage2023blind}. While traditional and emerging computer vision systems are typically celebrated for high accuracy rates, their accuracy rates plummet when used by blind people \cite{guo2020toward, gurari2018vizwiz,massiceti2023explaining}. Furthermore, AI VAT may be biased since common object recognition datasets are primarily trained on objects consumed and produced in the West \cite{nwatu2023bridging,de2019does, shankar2017no}, and OCR performs worse when processing non-English documents \cite{goyal2022flores,alginahi2017document}. For instance, the accuracy of commercial OCR in Arabic is estimated to be less than 75\% for printed text \cite{alghyaline2023arabic}, whereas it is 99\% accurate for printed English text \cite{chaudhuri2017optical}. 

Responding to calls to include disabled people as experts and co-designers of AI technologies, scholars have begun taking disability-centered approaches \cite{theodorou2021disability, kamikubo2023contributing, alharbi2022understanding, herskovitz2023hacking} to designing AI VAT. For example, Theodorou et al. engaged blind communities in the data collection process \cite{theodorou2021disability}, which led to increased representation of blind people in the datasets used to train AI systems. We add to this legacy by conducting an in-depth qualitative study with blind people to unpack errors they have experienced in AI VAT (e.g., processing limitations and cross-cultural bias) and highlight their perspectives for improving future AI VAT.

\subsection{Understanding \& Verifying AI Results}
AI systems are highly opaque and stochastic, leading to potential misalignment between users' expectations and systems' functionality \cite{eslami2016first, zhang2021ideal}. Users may have difficulty understanding AI outputs \cite{eslami2019user} which can cause misuse \cite{rossi2018building}. To bridge these gaps, HCI research has begun unpacking how users interpret and understand AI in various domains such as social media algorithms \cite{eslami2016first,simpson2021you}, healthcare \cite{cai2019human} and creativity \cite{chang2023prompt}.  One promising area that aims to support user understanding of outputs is explainable AI (XAI) \cite{miller2019explanation,lipton2018mythos}, a key facet of responsible AI efforts \cite{arrieta2020explainable, lee2020human, amershi2019guidelines}. However, there is no one definition of XAI. It broadly involves system transparency \cite{gilpin2018explaining, lipton2018mythos} and post hoc explanations of specific model outputs by displaying heatmaps \cite{maehigashi2023experimental, shitole2021one, kim2023humans}, communicating uncertainty \cite{zhang2020effect}, providing factual reasoning for why a specific prediction was generated \cite{ribeiro2016should,vstrumbelj2014explaining, ribeiro2018anchors} and explaining counterfactual cases \cite{miller2019explanation, keane2021if}. Previous empirical studies have demonstrated that users negotiate when and how to apply AI explanations in their decision-making process \cite{lai2019human,riveiro2021s}, and their perceptions of AI explanations often depend on their technical backgrounds \cite{ehsan2021explainable}. In addition to XAI, researchers have also called to empower users by enabling them to contest and challenge AI results \cite{lyons2021conceptualising}.

Much of the work on AI explainability has focused on non-disabled people. Despite disabled people being early adopters of AI \cite{bigham2018learning}, we know considerably less about how disabled people make sense of AI uncertainty, and what might explainability add to (or distract from) their existing process. This is crucial considering disabled people may not always be able to identify AI errors \cite{alharbi2022understanding, glazko2023autoethnographic}, leading to overtrust in AI systems \cite{macleod2017understanding}. A recent thread of accessibility work started to investigate how disabled people detect and negotiate AI errors. Huang et al. demonstrated how Deaf and hard of hearing (DHH) people detect false positives in sound recognition systems by validating with trusted hearing people \cite{huang2023not}. They argued that soliciting and incorporating user feedback in sound recognition systems empowered DHH people,  providing a greater sense of agency. In the context of visual access for blind communities, Abdolrahmani et al. studied blind people's reception to AI errors when navigating indoors \cite{abdolrahmani2017embracing}. They found that blind people's acceptance or rejection of AI error is highly contextual: errors that carried low risks were acceptable, whereas errors that may lead to socially stigmatized consequences were not (e.g., misidentification of bathroom gender signs). To mitigate these risks, Abdolrahmani et al. noted blind people would inquire sighted people and use mobility skills. Accessibility scholarship also studied error in the context of image description and alternative text \cite{macleod2017understanding,zhao2017effect,zhao2018face,wu2017automatic,gonzalez2024investigating}. Blind people worried about potentially posting embarrassing photos on social media, and they identified various information needs that should be incorporated in automatic alterntative text such as photo quality and key visual elements to help decrease risks \cite{zhao2017effect}. In formative work on the inaccuracy of alternative text, MacLeod et al. argued for designing for mistrust and communicating AI uncertainty through showcasing confidence ratings to blind people \cite{macleod2017understanding}. Recently, Gonzalez et al. conducted a diary study of an AI-powered scene description prototype to understand possible use cases and satisfaction with the technology \cite{gonzalez2024investigating}; they found that people generally lacked trust in its descriptions. They described that blind participants would sometimes test the AI with known images, indicating that blind people likely already have specific verification strategies that they use in practice. Building on this work, we aim to detail these tactics further, contributing an understanding of why and how blind people detect and verify errors in a variety of commercial AI VAT.

We join this relatively new line of research focused on accessible AI verification and explanations \cite{macleod2017understanding, abdolrahmani2017embracing,glazko2023autoethnographic} by foregrounding blind people's everyday verification experiences of AI VAT, and their desires for XAI features that better communicate AI output.

\section{Method}
We take a qualitative research approach to explore how blind people verify AI VAT errors, and the limitations and possibilities of emerging explainability features. We first requested that participants share three examples of using AI VAT with us. Drawing from these example scenarios, we conducted semi-structured interviews focusing on participants' current experiences with AI VAT, how they navigate uncertainty, and their desires for upcoming features. In what follows, we detail our recruitment strategy and procedure.

\subsection{Participation}
We collaborated with the National Federation for the Blind (NFB) to recruit adult participants who are blind or low vision and use AI VAT. We used a recruitment survey to confirm eligibility (at least 18 years old, live in the United States, and have used AI VAT). The survey asked respondents about their preferred visual assistance technologies, age, race, and gender to ensure diversity in our sample. Over 300 participants completed the survey, and we contacted approximately 50 respondents to schedule interviews.  We tried to reach out to trans/non-binary people, people of color, and older adults as their perspectives are often marginalized in HCI and accessibility research \cite{harrington2023working, sum2022dreaming, brewer2017xpress}. 26 out of the 50 respondents scheduled and completed an interview. We compensated participants with a \$35 (USD) Amazon gift card. This study was approved by our Institutional Review Board (IRB). 

All participants were daily or weekly users of VAT (with Seeing AI, Be My Eyes, and Aira being the most commonly used). Table \ref{tab:table1} provides a breakdown of VAT types used by participants. To preserve the anonymity of participants, we report the demographic information of visual disability, gender, and age in an aggregated format. In terms of visual disability, one participant had low vision, while the remaining $(n = 25)$ were totally blind with a mix of those who had light perception $(n = 12)$ and no light perception $(n = 13)$. Fifteen participants are visually disabled since birth, six acquired during childhood or adolescence, and five acquired during adulthood. In terms of racial/ethnic identity, ten participants are White, five are Hispanic/Latinx, four are Asian, two are Middle Eastern, one participant is African American and Native American, and one participant is Hispanic and White. Three participants did not report their racial or ethnic information. Fifteen participants identified as women and eleven identified as men. We collected participants' age range (e.g., 18-24, 25-34, etc), and the weighted average of the twenty five participants (one participant did not report) was approximately 41 years old. Participants worked or were interested in diverse fields such as technology, education, business, and art. Table \ref{tab:table2} in the appendix describes their technical background. 

\begin{table*}[ht]
\renewcommand{\arraystretch}{1} 
\centering
\begin{tabular}{wc{0.12\linewidth}|wc{0.11\linewidth}|wc{0.37\linewidth}|wc{0.04\linewidth}|wc{0.24\linewidth}}
\textbf{Type} & \textbf{Name} &  \textbf{Description} & \textbf{Count} & \textbf{Participant IDs} \\ \specialrule{.1em}{.05em}{.05em} 

\multirow{4}{4pc}{\makecell{Human \\ assistance}} 
    & \makecell{Be My Eyes} & Mobile, volunteer-based human assistance & 22  & \makecell{P1, P2, P3, P5, P6, P7, P9, P10, \\ P11, P12, P14, P15, P16,  P17, \\ P19, P20, P21, P22, P23, P24}          \\ \cline{2-5}
    & Aira & Mobile and desktop paid human assistance & 16  & \makecell{P1, P3, P6, P7, P8, P9, P10, \\ P12, P15, P16, P17, P18, P19, \\ P23, P24, P25}                  \\ \hline
    
\multirow{1}{*}{LLM-based AI} 
    & Be My AI & \makecell{Image descriptions from OpenAI's GPT-4, \\ with chat for follow-up questions} & 7  & P1, P4, P7, P8, P19, P20, P23          \\ \hline
    
\multirow{7}{4.5pc}{Traditional AI}        
    & Seeing AI & \makecell{Mobile computer vision for reading text, \\ recognizing color and light, \\ and describing scenes} & 26 & All participants \\ \cline{2-5}  
    & TapTapSee & Mobile computer vision for object detection & 13  & \makecell{P1, P2, P3, P4, P5, P6, P10, \\ P11, P15, P16, P17, P19, P26}                  \\ \cline{2-5} 
    & Envision AI & \makecell{Mobile computer vision for reading text and \\ recognizing objects} & 9  & \makecell{P1, P6, P7, P10, P11, P12, \\ P14, P15, P24}                                    \\ \cline{2-5} 
    & \makecell{KNFB Reader} & \makecell{Mobile OCR with text-to-speech, \\ text-to-Braille, and text highlighting} & 10  & \makecell{P4, P6, P7, P12, P14, P15, P18, \\ P20, P21, P24}                            \\  
\end{tabular}
\vspace{.5pc}
\caption{Summary and description of visual assistance technologies (VAT) used by our participants at the time of the study.}
\vspace{-1.5pc}
\label{tab:table1}
\end{table*}

\subsection{Procedure}

\subsubsection{Pre-interview scenarios}

Before conducting interviews, we asked participants to send us three examples of their AI VAT usage (e.g., Seeing AI, TapTapSee, and Be My AI). Specifically, we requested these examples to be around their confidence level (high, unsure, low) of the accuracy of AI output. Participants described recent or past interactions with AI VAT in written format or provided screenshots showing the output.

As we collected pre-interview scenarios for a couple of participants, we realized the potential of exposing participants to privacy risks and the additional work we might impose on participants to verify privacy leaks \cite{alharbi2022understanding, stangl2022privacy}. We also recognized that blind people may use AI VAT for sensitive contexts despite privacy risks to address access barriers \cite{alharbi2022understanding, stangl2022privacy}. We brainstormed an amendment that mitigates privacy risks while acknowledging the complexity of our request. After careful consideration, we used the following stipulation: \textit{``In case you are planning to actively use visual assistance technologies for the sole purpose of documenting examples for this study, please do not use sensitive information such as your ID, passport, credit card, social security card, or medical information.''} We hoped that this request would ensure that we did not expose participants to additional privacy harms while refraining from influencing how participants use VAT beyond this study.

\subsubsection{Semi-structured interview}
The first author led semi-structured interviews with blind and low vision people remotely over Zoom. Interviews were conducted from September to October 2023 and took approximately 45-60 minutes to complete. The interviews focused on various topics, including the promise and limits of AI VAT, why blind people may sometimes prefer working with sighted people instead of (or in addition to) AI, how blind people make sense of AI outputs, and what participants desired for future AI VAT. We updated the interview questions based on the pre-interview scenarios that participants shared. For example, we introduced follow-up questions on which applications participants used for different scenarios, why they thought the AI produced a particular output, and what a more appropriate output would ideally look or feel like. Interview questions can be found in the appendix. Each interview was audio-recorded and transcribed by the second author. 

\subsubsection{Data analysis} We followed a reflexive thematic approach \cite{braun2006using,braun2019reflecting} to analyze data. After one month of closely reading transcripts and open coding, early patterns included categorizing errors based on AI techniques (e.g., OCR or object detection) and surfacing blind people's error detection strategies. 
The first two authors refined these patterns by continuously (re)reading transcripts, writing memos, affinity diagramming, and having weekly discussions with other co-authors. We arrived at themes related to common errors blind people navigate when using AI VAT, strategies blind people use to make sense of these errors, and how they imagine future technologies to support their process of detecting and verifying AI errors. To further refine our early themes, we repeatedly returned to transcripts and searched for quotes that support, complicate, or extend these findings. In all parts of our research process, but perhaps especially during data analysis, we were aware of our positionality as sighted accessibility researchers. We acknowledge this as a limitation given our study's focus on the blind and low vision experience.

\section{Findings}
Our findings are organized into three sections around blind people's experiences with AI VAT outputs. First, we identify errors blind people encounter when using AI VAT. Then, we describe how blind people cultivate an intuition for assessing AI accuracy through 1) everyday experimentation with AI VAT in low-risk and known settings, 2) employing non-visual sensemaking skills, 3) collaborating with sighted bystanders and community members, and 4) cross-referencing various technologies and applications. Lastly, to support their verification and error detection strategies, participants highlighted opportunities for emerging explainability and contestability features within AI VAT. Taken together, findings unpack the technological, social, and cultural facets of navigating visual access.

\subsection{Errors in Visual Assistance Technologies} \label{FindingSec: Error}

In this section, participants identified common errors when using AI VAT: formatting and processing errors, as well as cross-cultural biases. We start by explaining blind people's encounters with errors when accessing complex real-world textual information, such as tables and blank lines designated for signatures. Next, we highlight cases of cross-cultural bias, noting how AI VAT fails to account for non-English languages and diverse cultural artifacts.

\subsubsection{Processing Errors} \label{Findingsub: process}

Participants described several errors they encountered in AI VAT such as color misrecongition and LLM confabulation. We will elaborate on these errors in later parts of our paper. Primarily, participants reflected on processing errors where AI VAT takes raw model output (e.g., labels/detections and bounding boxes) and processes that information to make it readable for users. Given limitations with the underlying models, as well as with how this data is processed, users sometimes experience errors.
For example, when using AI VAT to read documents, participants discussed uncertainties around how AI processes layouts such as date formats, tables, and blank spaces. P2, who sometimes uses Seeing AI to read work documents, described how the text `10/01' was processed as \textit{```You need to complete this paperwork by October 1st,' is what Seeing AI would say. When in actuality, the date was January 10th.''} AI VAT encoded the date and communicated an incorrect date format (i.e., instead of following the month/day format, it described day/month). Without knowing the raw text that was recognized, this result may seem reasonable. The ambiguity around AI processing is amplified when trying to read tables. P4 explained \textit{``take the example of this class schedule, right? So you take a picture with the document feature in the Seeing AI app. Somehow it does not recognize or it messes up because it's in a table format [...] you have to keep wondering, okay, this class and this class with so and so professor in so and so class.''} These errors may occur because AI VAT does not process table contents in a manner that is accessible and usable for blind people. In a similar frustration with how Seeing AI processes table content, P12 tried following a recipe on the back of a brownie box container. However, Seeing AI's document feature would read \textit{``a little bit of the nutrition information and a couple of lines of the recipe and then some more nutrition stuff and then some more of the recipe''} (P12). This fragmented reading was likely because of how the text is processed. Detected characters are put into an order prioritizing text on the same horizontal line, which will logically order a single-column document, but fail for many of the real-world tasks that blind participants wanted to complete.
Additionally, Seeing AI does not indicate non-text visual information on paper documents, such as blank input spaces. Some participants discussed this erasure as a form of error since AI VAT did not accurately convey important details of the document. P2 explained \textit{``if there's a checkbox, it'll read the information for the checkbox, it won't tell me that there's a checkbox. If there's a question, it'll read the information of the question. And then move on to the next question without letting me know there's a blank there for information input.''} Our findings emphasize that it is important to consider what textual, visual, and layout characteristics (e.g., blank space, table format) are valuable to blind people, and how to accessibly convey such information.  

\subsubsection{Cross-Cultural Bias} \label{Findingsub: cultural}
Some participants described OCR performance on non-English documents as a \textit{``nightmare''} (P14). When trying to use Seeing AI and Envision AI on Korean text, P14 told us \textit{``I couldn't read for nothing.''} AI VAT also failed to support some languages in the Global South as P10 explained \textit{``my native language, it is hard to come by. The Khmer language, the Cambodian language, it's not one of those languages that a lot of them [AI VAT] focus on.''} The lack of multilingual support could also impede accessing English content. P2 noted OCR did not work when trying to understand an English-Arabic translation of the Quran, the Holy Book for Muslims. They said \textit{``[Seeing AI] didn't know what to do with the Arabic. I think it tried to interpret it as random letters or even a picture of some sort. And then the English, because it was side-by-side, I think the Arabic confused it to the point where it didn't capture the English.''} While Seeing AI unfortunately has yet to support Arabic, P2 was surprised to learn that the application's OCR would attempt to decipher Arabic text as English. Beyond language and text-related errors, some participants mentioned that Seeing AI's Product does not recognize products that are less common in the U.S. P19 told us:

\begin{quote}
    \textit{``I have a Mexican store right next to me. It's a little store. A mom-and-pop shop that imports a lot of stuff and they have a lot of unique delicacies that you can't really find at Walmart. Ideally, I want to walk into this store and scan my way through, finding the snacks or the products that I need. But unfortunately, there have been situations where I buy some kind [of] soda or a product. I scan it and it takes me like six minutes to try to find the darn barcode and then when I find the barcode and [Seeing AI] takes its time to scan it, and it says: `Sorry, Seeing AI cannot recognize this, please try again.' And I'm like, what the heck? I'm just going to throw my phone out the window [...] So that's just incredibly frustrating and inequitable.''}
\end{quote}

Cross-cultural errors also occur when AI VAT described images and scenery. One participant used Seeing AI's scene feature to identify their graphic tee, and it resulted in a culturally offensive misrecognition where cultural dress was instead labeled as an animal. P10 explained \textit{``[t]he picture is supposed to be a woman. It's from my culture, [related to a] dance. It's called an Apsara. Instead of describing that the lady had a crown on, [Seeing AI described] it as an animal and [in my mind] I said: `No, I don't have a picture of an animal on that shirt.' I have a couple of graphic tees, but I know I don't have a picture of an animal on any of my shirts.''} Overall, findings affirm prior work that demonstrated how AI technologies, in general, tend to have a Western bias \cite{nwatu2023bridging,de2019does, shankar2017no, goyal2022flores, alginahi2017document}, and that AI-enabled assistive technologies often neglect to recognize the needs of blind communities from various racial and ethnic backgrounds \cite{alharbi2022understanding, bennett2021s}. 

\subsection{Process of Verifying AI Results} \label{FindingSec: Verify}

In this section, we detail how blind people verified AI VAT output. We start by noting that verification skills are developed through routine use. Then, we delve into more concrete strategies such as sensing objects as a means of verification, testing with sighted people, and switching between different devices and applications. 

\subsubsection{Verification Through Everyday Experimentation} \label{Findingsub: Everyday}

Blind people described verification as an orientation they built through their experience of using AI VAT. While AI VAT is commonly advertised as `seeing' for blind people \cite{sadjo2021landscape, bennett2020point}, P3 rejected this premise and affirmed that VAT is often wrong. P3 explained: 
\begin{quote}
   \textit{``If you think that seeing is this objective thing as opposed to interpreting, then that's a flaw in itself, and it is imbued in these [VAT] apps, right? Like, oh, we're seeing this for you because you can't, and so we're going to tell you what it is. And I think as a blind person, at least for me and some of the folks that I know, we know some of this is wrong.''}
\end{quote}

Challenging the misconception of equating AI VAT to normative sight, using these applications requires building familiarity and constant negotiation when interpreting results. 
As such, when we asked participants what advice they would give to blind people who are just starting to use AI VAT, most participants emphasized exploring these applications, noting that \textit{``[AI VAT] are fun. Just play with it. Don't rely on them. Find their strengths and weaknesses''} (P20). P24 added \textit{``make sure that you're not testing it on something that is going to impact your life in a negative way if the information you get is incorrect.''}

Through experimenting with AI VAT, participants discussed how AI VAT is often used for non-visual tasks where they already have a sense of what the correct information should be or they could easily verify. Reflecting on their experience, P7 said \textit{``I'm still very much kind of testing for myself what I can rely on. So, I've been very careful. I mostly use [AI VAT] in low-risk situations where either I already have the data [...] I've been a little bit more reluctant to completely rely on it for my first and only source of information.''} For example, P7 shared with us an incident where they dismissed AI VAT output. They used Be My AI in a bar they frequent and know very well. However, Be My AI produced incorrect results. P7 said  \textit{``[Be My AI described brands] I knew for a fact that they didn't sell.''} In trying to reason why this error occurred, P7 illustrated the logic of Be My AI as \textit{``if there's a tequila, there must be another tequila next to it. And I can't tell which tequila it is, so let me just make one up.''} This approach of exploratory use (i.e., testing AI VAT in known contexts) constructed a perception of AI VAT as \textit{``more of a guidance thing than a full support thing''} (P26). AI VAT becomes a mean to get surface-level information rather than details, and this critical use is shaped by prior experience. 

Participants avoided using AI VAT for visual tasks that are difficult to verify, and have a high probability of error. For example, \textit{``tasks such as maybe finding something that's been dropped or something more complex such as descriptions of pictures or looking at clothing to see if there are any stains''} (P1). While participants described some tasks (e.g., reading a document) as frequently containing errors, they still used AI VAT since such errors are easily detected and resolved by context. These errors were referred to as \textit{``classic OCR mistakes''} (P11) which involved character misrecognition. P12 explained that AI VAT would scan the letter {\fontfamily{qcr}\selectfont m} as the letters {\fontfamily{qcr}\selectfont rn}, noting \textit{``[OCR] does that often with web addresses. It will say `dot corn,' and I know it's `dot com.'''}  Similarly, P25 added \textit{``I've never heard of Heetos; I'm pretty sure that meant Cheetos.''} Because of the frequency of OCR errors, some participants mentioned that they have grown accustomed to \textit{``reading between the lines''} (P16) when using OCR. P12 explained \textit{``I'm so used to accounting for it without even thinking about it. It's like second nature to me.''} In essence, participants do not always need to verify every AI result, especially in cases where they have built familiarity and can correct errors on their own. Overall, in the case of AI VAT, blind people were critically aware of limitations, and that in turn shaped their skepticism and orientation toward verification based on use cases.

\subsubsection{Sensory Verification} \label{Findingsub: sensory}
Some participants employed non-visual senses (such as feeling and hearing) as a tool to verify AI outcomes. P16 said \textit{``I can shake a can of beans and a can of corn and can of green beans, or tomatoes, and I can kind of tell by the sound they make.''} Some participants described their tactile recognition technique as \textit{``usually when I touch something, I immediately know what it is''} (P13). Tactile recall depends on prior experience with the object of interest. P19 said \textit{``muscle memory of packages or bottles that I know that I've used.''} For example, P11 told us that one time an AI VAT described an item as \textit{``possibly ice pop or possibly ice cream bar.''} However, P11 said it was \textit{``a 3 pack of the corn on the cob that you get in the supermarket. Now I knew that was corn on the cob by feeling the outside of the package. It had cellophane around it just like a pack of meat.''} In the context of verifying a clothing's color, P15 told us \textit{``I was trying to find a particular dress shirt. [Seeing AI] was saying that it was light blue, but I was pretty sure by the material that it was the lavender one.''} By touching the material of their shirts and finding distinctive patterns (e.g., logos), some participants were able to refute AI's color recognition.

\subsubsection{Verifying \& Testing With Sighted People} \label{Findingsub: sighted}

Blind people sometimes involved sighted people to \textit{``figure out where the limits of AI and what can [they] feel pretty confident about''} (P25). For example, P25 verified some AI output with their sighted partner. Through their interaction, P25 was able to understand what types of content work well with OCR and what does not, concluding that OCR does not work well with \textit{``more colorful [packaging]''} (P25). Some participants explained making plans to reach out to trusted sighted people (e.g., friends or family members) in high-risk situations where accuracy and security are needed. P23 said \textit{``if I'm really kind of twinging that something's not right and it's something important, I will go seek out sighted assistance.''} Sighted people may also play a role in mitigating security leaks by helping with precise camera aiming before using AI VAT. P4 explained \textit{``I have to check how much I have to pay for my utility bill. So, in that case, I have asked my dad where exactly should I point my camera because it's always gonna be the same on that page without it revealing like my account number and everything.''} In low-risk settings, participants also described sometimes validating with sighted people when convenient. For example, P19 used Be My AI to get a sense of their staff room, and it described a vending machine. This felt weird to P19 since they \textit{``worked in different schools, in different settings, and have not encountered vending machines in the staff lounge.''} P19 confirmed when \textit{``somebody just opened the door and they're having conversations with other teachers and when they're done I asked `Hey, is there a vending machine?'''} Through this quick and mundane interaction, P19 verified that there was indeed a vending machine. Similarly, P6 verified AI VAT's color recognition with sighted people as they proceeded with other activities (e.g., en route to university). They said \textit{``if I'm getting picked up by paratransit, it is a service that takes people with disabilities to certain locations, I'll be like, `hey, what color is my shirt? TapTapSee or Seeing AI said this one color, I'm still a little skeptical.' [...] I'll periodically ask just to validate my inquiry.''} Corroborating prior work, we found that blind people sometimes verify results with sighted people \cite{zhao2018face}. We described how our participants confirmed AI outcomes with sighted people as a way to gain a better understanding of the limitations of AI. They also deliberately engaged with sighted people for high-risk tasks that require accuracy and security, whereas low-risk scenarios were verified spontaneously with sighted people as-needed.

\subsubsection{Cross-Referencing With Different Devices \& Applications} \label{Findingsub: devices/applications}
Blind people discussed switching between different devices and applications to verify AI results. In particular, they verified AI outputs by trying to confirm consistency. For example, some participants told us they would cross reference with \textit{``old school''} (P17) technologies like portable scanners. P4 said \textit{``I get a lot of letters from my college, so I skim those [using Seeing AI] [...] I also have a portable scanner. I do not know the name of the scanner, I scan through that also.''} One participant explicitly hoped that VAT applications would support blind people's reflection and verification process. P8 wanted a way to save and \textit{``consolidate the information in the chat feature [of Be My AI]''} so that they \textit{``could go back and compare [and] have that as a reference point.''} Participants also switched between different features within one application. P2 tried to read a paper using Seeing AI's Document mode. However, it produced an error message of \textit{``No image visible,''} (P2). To verify this issue, P2 \textit{``switched to Short Text [another feature in Seeing AI], then it recognizes text.''} Nevertheless, it is difficult to read long forms of text using Seeing AI's Short Text mode since users have to keep the camera stable otherwise it would \textit{``just start reading from the top again''} (P21).

While switching between different devices or applications gave blind people an avenue to validate AI, it did not always lead to a resolution. P26 took a photo of a room to get a sense of what was around them. TapTapSee and Seeing AI produced conflicting responses. They explained \textit{``the uncertainty came from the different wording in the descriptions like clock versus hat. Is it a clock? Is it a hat? Is it even really one of those things? [...] the specifics is where it got a little bit different. It didn't always overlap and so that's where the uncertainty came from.''} Similarly, P1 was \textit{``looking for a particular setting''} in their grill. P1 already had a sense of what the output should be, but they wanted to double-check using Be My AI. However, P1 said \textit{``when [Be My AI] gave me the order [of buttons], I could tell from that description that was not correct at all. There were buttons that were out of order.''} P1 could not cross-check with Seeing AI \textit{``because those are symbols that are on those buttons, not actual texts. Seeing AI would not have read those very well. It's designed to read actual text.''}

Furthermore, participants may sometimes cross-reference in creative ways, using devices that are not assistive technologies. P19 used Be My AI to read their work phone's extension number so they could \textit{``give my family this number in case of emergency.''} After receiving the extension number from Be My AI, P19 told us \textit{``I called this number because I just wanted to verify. If my office phone rings, great. If it doesn't, I don't know what's going on.[...] I got straight through the school counselor's phone line and I was like `okay, no, this is not it.'''} P26 added that when they use Seeing AI Product mode and it outputs an unfamiliar brand, they search for the brand online. They explained \textit{``if it gives me a brand that I've never heard of and I don't remember buying this. I can always type it in online and verify that it is in fact a brand for this product''} (P26).

\subsection{Blind People's Perspectives on AI Explainability \& Contestability} \label{findingsec: XAI}

Participants explained that it's inaccessible to non-visually interpret the images captured and processed by AI VAT, broadening XAI efforts beyond AI outputs. Then, participants discussed how XAI approaches, such as conveying confidence ratings in AI VAT outputs, may hinder and support their verification process. Lastly, participants articulated that AI VAT should incorporate avenues to challenge and contest AI VAT output.

\subsubsection{Interpreting Input by Interactive Camera Guidance} \label{Findingsub: AI input}

Participants discussed how errors and uncertainties may emerge because it is inaccessible to know what was inputted into AI VAT. As P23 explained, \textit{``sometimes I wonder whether or not the inaccuracy is really more my picture taking skills than actual inaccuracy.''} Blind people may not know what was captured by their camera \cite{jayant2011supporting, adams2013qualitative, chiu2020assessing} and processed by AI. Participants thought that some errors could be \textit{``user error, [that is] not pointing the camera properly''} (P16). For example, participants said their images could be blurry or too close to the object, resulting in faulty AI VAT output. Past work also demonstrated how the lack of camera guidance techniques can be inaccessible to blind people, and could lead to inaccurate AI output \cite{zhao2018face,hong2022blind}. Additionally, some images \textit{``can be very difficult to read. [...] if you have something neon, and it's got white text that could be very difficult to read for anybody whether sighted person or computer''} (P18). Accordingly, participants desired affordances that enabled them to understand image quality as it impacts AI performance. Some Be My AI users mentioned that this type of feedback is sometimes given. P4 said \textit{``[Be My AI] also tells you if I didn't take like the picture properly, or if there is an obstacle in that picture.''} However,  participants hoped that Be My AI would provide feedback in real-time (as a user is taking the photo), and suggest how to better retake the photo. P1 explained \textit{``if there was a way to tell the user how to improve the picture in future.''} 

Some AI VAT such as Seeing AI and KNFB Reader (now known as OneStep Reader) have built-in support to help guide blind people in capturing specific areas of interest. However, these current features are insufficient. P23 described existing guidance techniques as in their \textit{``infancy stages.''} P12 added \textit{``the [guidance] technology is there, but it's far from perfect.''} Overall, current camera guidance features are thought to be \textit{``sometimes helpful, sometimes it's kind of a pain in the butt. Not because it's wrong... They're just very visual''} (P25). For instance, participants found Seeing AI's guidance in Document mode particularly frustrating since it only indicates the document edges, leaving blind people to guess which direction they should move to correct the view. P10 explained \textit{``[Seeing AI] doesn't say move to the left or right. You kind of have to guess. If it says top left corner, that means your camera's kind of leaning more towards the left, so you kind of move it a little back to the right.''} Besides the perpetual need to remember to move in the opposite direction (i.e., going right when it indicates left edge detected), Seeing AI's guidance can be inaccessible because it fails to specify how much users need to move. P25 explained, \textit{``it feels like a game. [...] I'm thinking there are all these degrees. It's not just upright left. It is not just super simple like that. To what degree do I go down? So, when [Seeing AI] will say `left edge visible.' I'll go down and it'll say `left edge not visible' and I'll go down and It'll say `right edge not visible' and I'm like, [imitates screaming] you know.''} Seeing AI's guidance is further difficult on a round object. P10 tried to use OCR on a prescription bottle and said \textit{``it is hard just because of the way the bottle's made because it's round so you have to move the bottle for it to read what it's saying.''} 

In imagining how future VAT could improve, participants discussed opportunities to further refine camera position guidance. P14 reflected on their experience with Aira, a human-enabled VAT, and how they wished the guidance of AI technologies was similarly \textit{``interactive.''} They said \textit{``I go to Aira, I want to do like a team viewer session with Aira. So I pointed my iPhone camera at my computer screen and I asked [the Aira Agent], `can you see the IDM password on the team viewer?' They tell me `oh, go more to the left, go more to the right. Closer to your screen.' That's a lot more helpful than just `Oh, right edge detected, left edge detected [Seeing AI's guidance]' [...] So yeah, I wish it was more interactive like that.''} While redesigning an entire guidance system that is akin to interacting with sighted agents may be ambitious, a simple change to Seeing AI's guidance could be \textit{``just focusing on where you want the camera to be instead of where you don't want it to be''} (P8), eliminating an added cognitive burden. Participants also wanted the guidance to be based on shape. For example, when trying to perform OCR on cylindrical objects (e.g., bottles), VAT \textit{``could say something like `rotate right' or something like that as it's reading. Because maybe it would sense that the words were cut off and then the object was like a cylinder''} (P21).

Overall, participants discussed how it might be difficult to verify AI VAT output because it is inaccessible to know what was captured by their cameras and processed by AI. While some AI VAT does incorporate camera guidance features, participants felt these existing cues were insufficient. They envisioned future camera guidance to more interactive and less cognitively burdensome.

\subsubsection{Tensions Around Communicating 
Accuracy} \label{Findingsub: AI output}

Participants wanted to understand why AI produced a specific response. While prior work proposed incorporating confidence ratings in automatic alternative text \cite{macleod2017understanding}, participants in our study had mixed opinions about the benefits of indicating uncertainty in AI VAT. Some participants felt it was important to convey a level of uncertainty through quantifiable measures or phrasings of output. For example, in the context of outlandish misrecognition, P9 used an AI VAT to describe their cat and the application \textit{``said `rattlesnake.' I was like, okay... That's pretty hilarious because obviously there's not a rattlesnake here!''} When asked whether it would be helpful to get a sense of why AI generated this description. They elaborated \textit{``so if [AI] said like. `Yeah, 10\% rattlesnake.' It would at least give me some sort of nuance to the fact that [...] I know this isn't a rattlesnake. But there's something about the color of her fur that maybe looks like a rattlesnake.''} In the context of using AI VAT to read documents, participants described how getting a sense of accuracy would save them time. P20 told us \textit{``The app can tell you, `well, this is like 80\% accurate or 90\% accurate' Then you will know, it's not it's not a good OCR, so you might need to go find someone to read it to you.''} Preferring less quantifiable indicators, P2 explored verbal descriptors and emphasized the importance of indicating where in the document that uncertainty lies. They said \textit{``I don't know what phrase could be used, `Uncertain,'  `Less certain,' or `Possible.' Something like that. But especially indicating where that text is on the document or at least having a marker for it.''} 

Overall, some participants argued that it is critical to convey uncertainty, especially in LLM-enabled VAT (e.g., Be My AI) which often has humanistic undertones that might encourage overtrusting the output. P23 explained, \textit{``I worry about the misinformation problem. In the sense that If it suddenly sounds human and if it's giving you all of this reputable sounding information. Are you going to take it more realistically because it sounds human rather than a robot?''} However, one participant thought that communicating accuracy is an unreliable measure since AI VAT does not have ground truth. P22 elaborated:
\begin{quote}
    \textit{``I don't see how that would be feasible or practical [...] because [AI] is relying on its own measurements. So unless it has some way to measure against the original document and compare the outcome of the picture and OCR it performed, there's really no way to calculate those statistics.''}
\end{quote}

Few participants did not have strong opinions about communicating uncertainty in AI VAT. Some of these participants felt that they take a utilitarian and task-based approach to AI. When they experience an error in AI VAT, they do not want to waste time trying to understand the issue. P25 explained \textit{``I guess it doesn't seem relevant for me. Either it's working or it's not working. I don't really need too much information about it. I got too much stuff going on.''} One participant noted how these additional explanations can be problematic, especially during moments of AI bias where other actions are more appropriate. P23 said:
\begin{quote}
     \textit{``There are a lot of experiments that have been done around like AI and determining individuals of different races and different ethnicities and how it pretty consistently gets certain things wrong. Using that as an example, I don't really care why [AI] got it wrong. I cared that it got it wrong. Fix it. Make it better. Make it right.''}
\end{quote}

\subsubsection{Contesting AI Output} \label{Findingsub: Contesting}
Participants felt that the existing AI VAT systems do not reflect their preferred way of negotiating visual access. P3 told us that AI systems do not allow for shared \textit{``dialogue or discourse.''} Reflecting on their process of building visual access with sighted people, P3 said: 
\begin{quote}
\textit{``Well, I think when you have a [sighted] person, whether it's via FaceTime or Aira or in person, you have a back and forth [conversation] between that person [...] So if somebody reads to me something and I have a form to complete, I'm not going to just let that person take over. It's going to be back and forth.''}
\end{quote}

P23 added \textit{``with a human not only are they going to give me a description, but I can interrogate that description, right? Okay, so you say it's blue. Well, is it dark blue? Is it light blue? I can refine.''} These quotes emphasize that blind people play a critical role in shaping access \cite{bennett2018interdependence} as demonstrated by participants' active negotiation with sighted people. Some AI VAT limit blind people's ability to contest its output (i.e., challenge AI systems \cite{almada2019human, lyons2021conceptualising}). As P20 explained, \textit{``you cannot really ask [Seeing AI] a question because you take your picture and you get the description and that's it. So, you cannot really interact with [Seeing AI].''} However, participants shared strategies they currently enact to contest and improve AI VAT output. They also articulated potential avenues for future VAT technologies.

Some participants used Be My AI's chat feature to contest and teach AI. Recalling the incident where Be My AI misrecognized their work phone's extension number (shared in \ref{Findingsub: devices/applications}), P19 said \textit{``I can also type in my feedback, say `Hey, this is not the extension number.' It comes back saying `sorry, I apologize for the mistake.' So I always like to do that because then as a user I have the power to improve it.''} When asked why they think it is important to provide feedback to Be My AI during chat sessions, P19 elaborated \textit{``I really like giving feedback to the AI-powered models more often because they're AIs, they're computers and they sometimes hallucinate and make mistakes. That's why I really like to give feedback to those types of apps than others because they have potential. They are our future and we can't deny it.''} There is a sense that by fine-tuning AI during chat sessions, AI systems would eventually ``learn'' and become better. However, that is not always the case. P20 used Be My AI to get a visual description of their photo on a nature trail. Be My AI incorrectly described their white cane as a \textit{``hiking pole''} (P20). They followed up by asking \textit{``is it a hiking pole or a blind cane or a white cane?''} Be My AI continued to describe their white cane as a hiking pole. P20 explained
\textit{``normally with the online platform you can tell ChatGPT the answer is not right. This is the correct answer. [ChatGPT] is not gonna argue with you, right? So it just says `okay, yeah' So I'm not sure yet how I can do that with [Be My AI].''} In this case, the AI VAT did not immediately improve its output even after an attempt at contesting.

Participants wanted accessible and usable interfaces for blind people to contest output and provide feedback. In imagining how to incorporate feedback opportunities in AI VAT, P21 said \textit{``just a little pop-up and it says, `how did we do?' and [users] would have the option to skip it if [they] didn't have time for that.''} One participant, P23, emphasized that feedback requests ought to be accessible and usable by blind people who are multiply disabled. P23 explained that VAT companies should be mindful that\textit{``blindness is not a single disability population [...] cross-disability perspectives are not a common thing in [...] typical blindness apps.''} Indeed, erasing the experience of multiply disabled people is unfortunately a common occurrence in assistive technologies \cite{mack2021we}. When designing feedback forms, P23 elaborated that developers need to consider questions like \textit{``what methods do you have for folks who don't type but who use like augmentative communication aids?''}

Some participants voiced concerns about providing feedback to AI systems. Primarily, participants worried about being constantly bombarded by feedback requests. Reflecting on Be My Eyes feedback prompt (thumbs up or down icons) after interacting with sighted volunteers, P12 said \textit{``when I end the call and I hit thumbs up, it's not a big deal, but I use Seeing AI so often that if every time I finished a scan or something, and it [requested feedback], it would drive me crazy.''} P12 explained that it is important to have frequent feedback prompts when interacting with sighted volunteers because it may keep volunteers accountable whereas it is unclear how to \textit{``to hold an automated app accountable.''} In part, this could be because some errors are indicative of complex issues such as \textit{``there's a bug in the app''} (P12), and it would be frustrating to repeatedly point out this problem. Instead, participants opted for less frequent feedback engagements. P9 felt that it is more meaningful to give direct feedback by conducting compensated interviews or surveys because writing feedback is work, and it should be valued as such. They asserted:
\begin{quote}
     \textit{``It's a lot of mental energy to have to go into the app and find the `give us feedback' thing. Then, I gotta get out my Bluetooth keyboard because it's gonna take forever to type something on my screen or I know that the voice dictation is gonna mess something up [...] pay me well to sit down and honestly give my time and energy, and I will.''}
\end{quote}

\section{Discussion} 
Our results reveal the errors blind people encounter with AI VAT, the strategies they use to confirm or reject AI outputs, and opportunities for supporting accessible XAI. We turn to the feminist disability framework of misfitting \cite{garland2011misfits}, a concept produced by disability studies scholar Rosemarie Garland-Thomson to rethink disability and access. Situated in feminist literature on vulnerability and dependence, Garland-Thomson employs the metaphors of ``misfitting'' and ``fitting'' to demonstrate how embodied experiences are constructed by material relationships between bodies and worlds, seemingly comfortable when there is a fit and discordant when there is a misfit. Misfitting and fitting arise because of dynamic interactions between disabled people and structures. However, misfitting is \textit{not} an inherent quality of disabled bodyminds. For example, Garland-Thomson recounts the experience of wheelchair users trying to access a building: they may fit when there are elevators, and misfit when there are only stairs. Misfitting, with the injustice and violence it often accompanies, spotlights disabled creativity and ingenuity. For Garland-Thomson, misfitting becomes a site to affirm disabled ways of knowing. 

We build from this theory to illuminate how misfitting unfolds in access technologies. In the context of AI VAT, misfitting refers to errors that arise either from technical structures or mismatches with how blind people use these systems. The concept of misfitting is an analytical lens that interrogates what it means to fit or misfit into a system. Misfitting builds on long-standing traditions in accessibility scholarship to critically examine assistive technologies. The overall goal is not to make marginalized people "fit" into technologies. Historically, that often encouraged predatory inclusion practices \cite{bennett2020point, taylor2019race}. Rather, the misfitting framework offers language that demonstrates how blind people creatively adjust their behaviors to verify inaccessible AI, while also calling into question the concept of ``fit'' and whether they should have to or want to do that verification labor. 

\subsection{Who `Fits' \& `Misfits' in AI Systems?}
The misfitting framework enables us to interrogate assumptions about fitting and misfitting in our world \cite{garland2011misfits}. We use the concept to argue that blind people experience misfitting in AI VAT as computer vision prioritizes sighted camera aiming. Blind people with marginalized identities and want to access non-English text and non-Western materials experience further misfitting.

\subsubsection{Misfitting in Computer Vision Systems, and the Inadequacy of Camera Guidance Techniques:} In the context of AI, foundational computer vision systems are often trained on sighted people's data, making them optimized for sighted camera aiming. When used by blind people, the accuracy of computer vision systems decreases \cite{gurari2018vizwiz,massiceti2023explaining}. In other words, there is often a \textit{fit} when sighted people use computer vision, and a \textit{misfit} when used by blind people. Consequently, AI VAT embedded camera guidance features in hopes of making blind ways of camera aiming more legible to AI VAT (i.e., addressing the misfit). However, we learned from our participants that these guidance features are far from adequate. Current camera guidance systems not only maintain the existing misfit but also contribute an additional layer by potentially centering sighted logic. As noted in section \ref{Findingsub: AI input}, the current guidance is not aligned with how blind people prefer to receive directional information (e.g., P25 reflected on how language is visually-centric, and P10 articulated the additional cognitive labor to make sense of where to orientate their camera based on cues). These communication breakdowns between participants and camera guidance cues could be because such features do not incorporate what blind sociologist Siegfried Saerberg coined as ``blind style of perception,'' denoting sensory schemes of interpretation that blind people develop in relation to spatial and embodied processes \cite{saerberg2010just}. Instead, current guidance cues might be based on sighted styles of perceptions, and this clash may occur because blind people cannot negotiate what type and level of guidance is more meaningful to them. A one-size-fits-all approach to camera guidance in AI VAT fails to capture the diverse ways blind people wish to receive directional camera cues based on their access needs in the moment. Our analysis offered preliminary insights into making these camera guidance cues more interactive and malleable based on object type (recall P21 and P10 notes on using AI VAT for round objects). We encourage future work to build from our study and prior work on blind photography in general \cite{jayant2011supporting, adams2013qualitative, chiu2020assessing} to design better camera guidance features for AI VAT. For instance, upcoming work could explore camera guidance techniques that can detect object shape, and provide directional cues with length indicators (e.g., specific measurements or verbal descriptors) or guidance on how much to rotate round objects.

\subsubsection{Marginalized Blind Communities Encounters With Misfitting:} Our analysis demonstrates how AI VAT may privilege the experiences of an imagined `typical' blind user, neglecting those who do not fit this ideal. In thinking about who is seen as a `fit' to mainstream society, Garland-Thomson asserts that to ``dominant subject positions such as male, white, or heterosexual, fitting is a comfortable and unremarkable majority experience'' \cite{garland2011misfits}. That is, experiences of misfitting are often racialized and gendered. Building from our findings and prior work \cite{harrington2023working, alharbi2022understanding, bennett2021s}, we argue that blind people who hold marginalized identities (e.g., being an ethnic minority in the US) misfit in general-purpose AI VAT and computer vision systems, as previously discussed.  AI VAT are often marketed as `international' and general purpose, attending to the needs of blind people around the world \cite{sadjo2021landscape}. However, our findings complicate this narrative. In section \ref{Findingsub: cultural}, participants noted instances of cross-cultural bias, articulating how AI VAT do not attend to their cultural products, artifacts, and languages. P19 shared their frustration of being unable to use Seeing AI in a local Mexican store. P10 felt that their native language, Khmer, is not of interest to AI VATs. These cultural and linguistic erasures may lead to multiple types of harm as described in Shelby et al.'s taxonomy of algorithmic harms \cite{shelby2022identifying}. For example, it may result in quality of service harms \cite{shelby2022identifying,bird2020fairlearn} because of performance discrepancies between blind people who speak English and prefer Western products and those who do not primarily speak English and consume non-Western foods. In the emerging research area of fair and just AI for disabled people \cite{bennett2020point, whittaker2019disability,newman2023definition}, upcoming work must challenge the tendency to think of disability as a singular identity and recognize how intersectionality \cite{harrington2023working, crenshaw1989demarginalizing, berne2018ten} critically shapes experiences of AI error.  What we, AI and accessibility researchers, define as universal or general purpose is likely removed from reality and perhaps reflective of Western norms \cite{hoffmann2021even}. Nevertheless, there is an exciting trend in research that aims to empower blind people in shaping personalized object recognition  \cite{ahmetovic2020recog, kacorri2017people}. We hope this translates to commercial AI VAT, in addition to improving cross-cultural OCR and image description.

\subsection{Misfitting Celebrates Disabled Ways of Knowing}
Garland-Thomson argues that disabled people, through their experience of misfitting, cultivate expertise in building access \cite{garland2011misfits}. By experiencing misfitting in a particular time and space, disabled people gain a sense of resourcefulness to circumvent inaccessibility. For example, she notes that blind people uniquely navigate the world without relying on vision, a skill sighted people lack. Specifically, blind epistemology, as Caroline Jones writes, ``demands a rethinking of how we form knowledge” \cite{jones2020global}. In turn, this shifts the tendency to frame blindness and disability as a ``problem'' to a source of creativity and a position of knowing.

Despite the lack of affordances to support verification in AI VAT, our findings emphasize the creative ways blind people confirmed or rejected AI VAT. Contrary to the dominant discourse around AI VAT as ``seeing” for blind people \cite{bennett2018interdependence,sadjo2021landscape}, participants challenged painting sight and AI as objective forms of truth (recall P3 quote in section \ref{Findingsub: Everyday}), affirming blind people's critical role in the process. By applying the misfitting commitment of emphasizing ``remediation over origin'' \cite{garland2011misfits}, our analysis reveals how blind people use complex and ordinary methods to make sense of AI when faced with error and uncertainty. In particular, findings described how blind people engaged in intrasubjective and embodied practices. For instance, in section \ref{Findingsub: sensory}, participants employed non-visual sensemaking, like feeling or hearing objects, as a means to dismiss AI VAT. Complementing Gonzalez et.'s research on scene description \cite{gonzalez2024investigating}, we learned that blind people test AI VAT in low-risk and known contexts to inform future use. Our participants also discussed their gained familiarity with reading text filled with \textit{``classic OCR mistakes”} (P11 in section \ref{Findingsub: Everyday}). Additionally, participants described intersubjective ways to verify AI, such as strategically engaging sighted people. Building from prior work on interdependence \cite{bennett2018interdependence, mingus2017access}, which emphasizes how access is co-created by both disabled and non-disabled people, our findings highlighted the mundane and deliberate ways blind people included sighted people to verify AI. Taking these strategies together, we emphasize that the verification process undertaken by blind people is not additive. Rather, it is a critical step that enables visual access.

Overall, failing to support blind people's existing verification efforts within AI VAT applications may potentially weaken genuine AI partnerships, a value that disabled communities hold as noted in our findings and prior work \cite{newman2023definition, herskovitz2023hacking, gadiraju2023wouldn}. A misfitting lens attunes us, accessibility and AI researchers, to the creative epistemic labor developed by disabled people. We advocate for designing AI VAT systems that are aligned and in harmony with blind people's verification process. For instance, P8, in section \ref{Findingsub: devices/applications}, proposed that AI VAT should allow users to save outputs to support blind people in cross-checking with other applications. Using our findings as a starting point, we invite researchers to co-design AI VAT verification affordances with blind communities. Future work could also empirically extend this line of inquiry by developing typologies with blind people on the types of AI uncertainties they face in various contexts and how their verification strategies may evolve as new models emerge.
\section{Lessons \& Directions for Responsible AI}

In this section, we outline specific takeaways for Responsible AI, an evolving domain in both research and industry that foregrounds principles like transparency, explainability, and inclusion \cite{arrieta2020explainable, lee2020human, amershi2019guidelines,wright2020comparative}, within AI VAT teams and beyond. Particularly, we call on Responsible AI practice to 1) prioritize accessible XAI, and 2) work toward disability-centered audits.  

\subsection{Extending and Moving Beyond Explaining AI Outputs}
Findings provided insights into blind people's preferences for framing AI explanations. Participants articulated how explainability may sometimes support their verification strategies (e.g., when using OCR, understanding how AI is processing complex layouts such as tables is useful). Our participants' accounts also affirm prior work on the limitations of confidence scores \cite{alharbi2022understanding, macleod2017understanding}. While some found confidence scores helpful in certain contexts such as getting a sense of the accuracy of long documents when using OCR, participants questioned the validity of these metrics since they were unsure of how accuracy is computed (as P22 explained in section \ref{Findingsub: AI output}). Indeed, Gilpin et al. argued that explainability alone is not enough; it ought to be coupled with the ability to articulate outputs, answer users questions, and be subjected to auditing \cite{gilpin2018explaining}. Future work could study whether factors such as describing how confidence scores are calculated would help blind people negotiate trust.

 Prior research on computer vision explainability may assume that users' interpretation of visual input is trivial or self-evident. Recently, Hong et al. explored the potential and limitations of providing quality descriptors (e.g., blurry image) in a teachable object recognition prototype for blind people \cite{hong2022blind}. They found that participants benefited from such information to further iterate and improve AI performance. Our study similarly demonstrates that blind people value understanding the quality of their images even in contexts beyond training object recognition (e.g., OCR). We argue that explaining AI input -- especially when coupled with accessible camera guidance -- serves as a site to support verification, potentially resolving uncertainty around the source of error. This empowers blind people to understand whether the error they encountered was due to poor image quality issues or model limitations. 
 
Building from human-AI design guidelines \cite{amershi2019guidelines,wright2020comparative}, our findings teach us that Responsible AI scholarship should advocate for 1) providing transparent explanations on how confidence ratings are produced (i.e., explaining the explanations), and 2) challenging ability assumptions \cite{wobbrock2011ability}; describing key attributes of the input that may be inaccessible to users.

\subsection{Supporting Disability-Centered Audits}
In imagining better futures, some participants envisioned more direct engagements with AI VAT. For instance, our participants emphasize the need to contest AI outputs through feedback. However, the majority of AI VAT are closed and proprietary which is a stark departure from its  participatory\footnote{In tracing the precursor and foundations of OCR technologies (known as optophone), scholars asserted that blind people were not merely testers of such technologies; they were also co-developers by offering recommendations of hardware and demonstrating how its used to the public \cite{chan2018optophonic, mills2015optophones}.} origins. \textit{What might it mean to include blind people's perspectives in the development of AI?} Recently, accessibility scholarship has suggested developing AI audits with disabled people to address harms \cite{gadiraju2023wouldn}. Our findings offer some insights into how blind people can be involved in improving AI, and pave the way toward disability-centered AI audits. Here, we call for flipping the script of designing AI to ``replace'' or ``extend'' blind people's abilities, and towards enabling blind people to inform AI \cite{bigham2018learning, thieme2020interpretability, bennett2020care}.  Some participants felt excited and even compelled to be actively involved in reshaping AI (recall P19 comment in section \ref{Findingsub: Contesting}). While some AI VAT are open to receiving feedback (e.g., Be My AI noted in their blog: ``[...] please be patient, and keep telling us about your experiences, positive and negative, so we can make this the best possible tool for you'' \cite{BeMyAIannouncement}), it is unclear where or how to provide feedback. Participants envisioned providing feedback on the spot (e.g., in Be My AI's chat), and through formal sessions and surveys. They also emphasized equitable compensation for their work in identifying errors, and accountability measures describing how their feedback would be implemented. One way to help blind people collectively contest and repair AI VAT could be through disability-centered audits that gather their experience on AI outputs that can be improved, and objects that require further AI training. Specifically, in designing these audits, AI VAT applications may draw inspiration from ACCESS SERVER \cite{MELT2023}, a design research project that anonymizes and compensates disabled people for finding access barriers in cultural institutions. Building from a legacy of disability activism and scholarship, ACCESS SERVER affirms the agency of disabled people, makes the feedback process more accessible by providing optional templates, and values their labor by financially compensating them. AI VAT can adopt such practice within applications by \textit{clearly} stating how their users data will be used and handled, compensating users for their data, and reporting on any changes as result of their data. Furthermore, our findings on cross-cultural bias assert the importance of reshifting our focus on quality and accuracy when auditing AI VAT, paying particular attention to cultural representation. 

\section{Limitations} We had several limitations related to recruitment. We tried to recruit a diverse sample of participants in terms of age. However, our current sample only included one older adult participant (i.e., over 65). Our findings may not capture the experience of blind and low vision older adults. While our sample did include participants who have diverse cultural backgrounds and speak multiple languages, our study is based in the U.S. and may not extend to the majority of the world. We also recruited participants at a critical stage in the VAT technology scene: Be My AI was only open to beta users, and we included some $(n = 7)$ participants who have used Be My AI. Other VATs (e.g., Seeing AI, TapTapSee, and Aira) recently incorporated LLM features after we concluded our study, which may impact blind people's future experiences with them. 

\section{Conclusion}
AI systems will never be 100\% accurate. While there have been substantial efforts to recognize and address AI errors, most of those efforts have ignored blind people, resulting in verification processes that are inaccessible to them. This qualitative study sheds light on blind people's use of visual assistance technologies to non-visually verify AI outputs. 
We found that blind people often experience errors when using AI VAT to read complex layouts, and we detail cases of cross-cultural bias. To verify AI VAT results, blind people employed various tactics such as experimenting in low-risk contexts, using non-visual sensemaking skills, strategically including sighted people, and cross-referencing with other devices and applications. To enhance AI VAT, blind people desired more interactive camera guidance to negotiate AI errors. Some participants complicated common XAI techniques such as confidence ratings, noting ambiguity around how these indicators are computed. Instead, they emphasized avenues to directly contest and improve AI outputs. Extending our findings, we applied the feminist disability framework of misfitting/fitting as a generative perspective. We argued that blind people ``misfit'' in computer vision systems whereas sighted people fit. Blind people who are racially or ethnically marginalized experience an additional layer of misfitting since AI VAT does not account for their lived experiences. Furthermore, we called attention to the creative ways blind people negotiate misfitting in AI VAT systems that often invisiblize their verification work. Finally, we offer provocations for the field of Responsible AI, underscoring the need to prioritize accessible XAI and work toward disability-centered audits.

\begin{acks}

This research would have not been possible without the time and expertise of our participants. Thank you to the 26 participants of this study for sharing your insights with us. We thank the National Federation of the Blind (NFB) for circulating our recruitment survey. We also appreciate Aashaka Desai, Andrea Jacoby, Cami Goray, Jessica Flores and Sam Ankenbauer for providing generous feedback and comments on earlier drafts. 

This material is based upon work supported by the National Science Foundation under Grant 1763297 and by a Google Research Inclusion Award. Any opinions, findings, and conclusions or recommendations expressed in this material are those of the author(s) and do not necessarily reflect the views of the National Science Foundation.

\end{acks}

\bibliographystyle{ACM-Reference-Format}
\bibliography{sample-base}

\appendix
\section{Interview Protocol}
\textit{Note to readers:} the following questions are merely an outline of the major topics we hoped to discuss with participants. Given the flexibility granted by semi-structured interviewing, we often deviated from this protocol, asked follow-up questions based on the specific stories that our participants shared, and tried to mimic participants' language as much as possible.
\subsection{Current VAT use:} 

 \be
    \item You mentioned in the recruitment survey that you use [insert types of VAT].  Are there any other applications you would like to add or remove to this list? 
    \item In general, how do you use [AI VAT]? 
    \item If you could give advice to a blind person who is just starting to use these computer/AI-enabled applications what would you say?
\ee

\subsection{Pre-interview scenarios}

In our email exchange, you shared with me three examples of using [AI VAT]. I am going to go over these examples and ask a couple of questions. 

\subsubsection{Example 1: Low confidence of AI}

For the first example, we asked you to share a scenario where you were confident that the visual assistance technology provided the wrong responses. You sent us [briefly describe this photo or read text] and [AI VAT] provided a response of [read].

\be
\item Can you tell me more about this example? Would you typically use [insert AI VAT name] for this task?
\item What makes you confident that [insert AI VAT name] had the wrong output?
\item Can you tell me your best guesses for why [insert AI VAT name] produced this particular result?

\item Would you say other applications like [mention other AI VAT participant uses] could provide more accurate results?

\item How do incidents like this, where VAT produced wrong responses, shape your experience of using this VAT in the future? 

\item In general, are there particular types of visual information or scenarios where you feel that VAT would fail to provide accurate responses? Can you share an example?

\item If any, how do you address cases when AI VAT produced an error? 

\ee

\subsubsection{Example 2: Medium confidence/unsure of AI}

For the second example, you shared a scenario where you were unsure if the visual assistance technology provided an accurate response. As a reminder, [briefly describe this photo or recap text] and [AI VAT] provided responses of [read].

\be

\item Can you tell me more about this example? What makes you unsure of [insert AI VAT name] output?

\item How did you resolve this uncertainty? 

\item If applicable: [read the response of AI VAT]. What do you think of this response? How does the phrasing of this part shape your confidence about its quality?

\item Beyond [discussed example], are there particular types of visual information or objects, that when using VAT, you often find yourself unsure of the credibility of its response? If so, tell me a recent example. 

\item Can you tell me about a time when you double-checked information from [insert name of AI VAT] using alternative sources maybe like another application, friend, or family member? What was the outcome? 

\ee

\subsubsection{Example 3: High confidence of AI}

So for the first example, you shared a picture where you were confident that the visual assistance technology provided correct responses. As a reminder, [briefly describe photo or read text] and [VAT] provided the response [read].

\be
\item Can you tell me more about this example? 

\item What makes you confident or sure of [insert name of VAT] output in this example?

\item Can you give me other recent examples where you were certain that [VAT] produced accurate information?

\item Can you tell me about a time when you were confident about the response you received from VAT, but you later learned that it might have been inaccurate or wrong?
 
\ee

\subsection{If the participant uses human-VAT: error in human VAT}

So far we've discussed your confidence in AI-generated responses. I'm going to ask you a couple of questions about your experience in human VAT. 

\be

\item Can you tell me about a time when you requested visual assistance from a volunteer/agent and you were uncertain about their response?

\item How often does this happen?

\item How is your process for accessing response quality different from when using humans vs. AI?

\ee

\subsection{Future technologies}

\be
\item In your opinion, what types of technology features AI VAT could introduce to help you better assess the quality or credibility of AI responses?

\item If the creators of AI VAT could explain its general process for how it produces responses, so for example, VAT would accessibly explain how their system works at a high level, do you think this may help you in understanding the quality of its particular response? Why or why not? 

\item If any, can you share an example of a situation where having an explanation for how a VAT response is generated would have been particularly useful to you in understanding its accuracy or credibility of AI VAT?

\item If any, what potential risks or harms could arise from these explanations? 

\newpage
\section{Participants Overview}
\begin{table}[h]
\begin{tabular}{l|l}
\textbf{P\#} & \textbf{Self-identified Technical Background}     \\
\hline
P1                          & Can solve most issues with some help from friends or \\
& family members or online \\
\hline
P2 & Can solve most issues but then ask help from friends \\
& or family members or online \\
\hline
P3                          & Could easily solve most or all issues you encounter by yourself               \\
\hline
P4                          & Can solve most issues with some help from friends \\
& or family members or online \\
\hline
P5                          & Can solve most issues with some help from friends \\
& or family members or online \\
\hline
P6                          & Seek professional help to fix technical issues                                \\
\hline
P7                          & Could easily solve most or all issues you encounter by yourself               \\
\hline
P8                          & Can solve most issues with some help from friends or \\
& family members or online \\
\hline
P9                          & Could easily solve most or all issues you encounter by yourself               \\
\hline P10                         & Can solve most issues with some help from friends \\
& or family members or online \\
\hline
P11                         & Could easily solve most or all issues you encounter by yourself               \\
\hline
P12                         & Could easily solve most or all issues you encounter by yourself               \\
\hline
P13                         & Could easily solve most or all issues you encounter by yourself               \\
\hline
P14                         & Can solve most issues with some help from friends \\
& or family members or online \\
\hline
P15                         & Could easily solve most or all issues you encounter by yourself               \\
\hline
P16                         & Can solve most issues with some help from friends \\
& or family members or online \\
\hline
P17                         & Can solve most issues with some help from friends \\
& or family members or online \\
\hline
P18                         & Could easily solve most or all issues you encounter by yourself               \\
\hline
P19                         & Can solve most issues with some help from friends \\
& or family members or online \\
\hline
P20                         & Ask friends and family members to fix issues                                  \\
\hline
P21                         & Can solve most issues with some help from friends \\
& or family members or online \\
\hline
P22                         & Can solve most issues with some help from friends \\
& or family members or online \\
\hline
P23                         & Could easily solve most or all issues you encounter by yourself               \\
\hline
P24                         & Could easily solve most or all issues you encounter by yourself               \\
\hline
P25                         & Can solve most issues with some help from friends \\
& or family members or online \\
\hline
P26                         & Could easily solve most or all issues you encounter by yourself  \\
\end{tabular}
\caption{Participants' response to the recruitment survey question of ``When it comes to solving technical issues, you often:'' with options: 1) ``Could easily solve most or all issues you encounter by yourself,'' 2) ``Can solve most issues with some help from friends or family members or online,'' 3) ``Ask friends and family members to fix issues,'', and 4) ``Seek professional help to fix technical issues.'' P2 clarified their response during our interview.}
\label{tab:table2}
\end{table}

\ee

\end{document}